\documentclass{article}  
\usepackage{breckenridge}
\usepackage{amsmath,amssymb,citesort,enumerate,here}

\newcommand{\eqeqref}[1]{Eq.~\eqref{#1}}
\newcommand{\eqseqref}[1]{Eqs.~\eqref{#1}}
\newcommand{\refref}[1]{Ref.~\cite{#1}}

\newcommand{\ds}{\displaystyle}

\newcommand{\Nmj}{N^{mj}}
\newcommand{\Nch}{N^{ch}}
\newcommand{\Npart}{N_{part}}
\newcommand{\phobos}{{\sc phobos\ }}

\frompage{000} \topage{000}                                              %|
\def\rightmark{Saturation as Glauber-Gribov multiscatterings}
\def\leftmark{A. Accardi et al.}
 
\title{An interpretation of saturation phenomena \\
 as Glauber-Gribov multiple parton scatterings}
\authors{
{A. Accardi and M. Gyulassy %
}\\[2.812mm]
{\normalsize
Columbia University, \\
538 West 120th Street, New York, NY 10027, USA\\[0.2ex] 
}}
 
\abstract{
We compare two formalism that describe minijet production in $pA$ and
$AA$ collisions: pQCD supplemented by Glauber-Gribov 
multiple semihard parton scatterings
(pQCD+Glauber), and the Colour Glass Condensate (CGC).
We argue that in a suitable limit they are
equivalent to each other, the PQCD+Glauber model being more 
accurate from a numerical point of view.
Finally, we analyze RHIC data on Au-Au integrated charged
multiplicities in
the pQCD+Glauber framework, and conclude that at least at central
rapidity there is no sign of gluon saturation.}

 
\begin{document}
 
\maketitle
\setcounter{page}{1}

%%%%%%%%%%%%%%%%%%%%%%%%%%%%%%%%%%%%%%%%%%%%%%%%%%%%%%%%%%%%%%%%%%%%%%%%%%
%%%%%%%%%%%%%%%%%%%%%%  SECTION 2     %%%%%%%%%%%%%%%%%%%%%%%%%%%%%%%%%%%%
%%%%%%%%%%%%%%%%%%%%%%%%%%%%%%%%%%%%%%%%%%%%%%%%%%%%%%%%%%%%%%%%%%%%%%%%%%

\section{pQCD, Glauber-Gribov parton rescatterings \\
and the colour dipole} 
\label{sec:pQCD}

The model of Ref.~\cite{AT01b}, hereafter labeled ``pQCD+Glauber'', 
assumes minijet production in $pA$ collisions to be dominated by
semihard parton multiple scatterings. By ``semihard scattering'' we
mean a process with a minimum transverse momentum exchange $p_0
\approx 1-2$ GeV, described by leading order (LO) pQCD parton-parton
cross section.
The model 
assumes the S-matrix for a collision of one parton 
on $n$ partons from the target to be 
factorizable in terms of S-matrices for parton-parton
elastic-scattering, and assumes generalized pQCD factorization.
Considering only gluons for simplicity, 
the minijet transverse spectrum is then given by
\begin{align}
    \frac{d\sigma^{mj}_{pA}}{d\eta d^2p_T} = 
        G(x,p_T^2) \frac{d\sigma^{\,A}_{hard}}{d^2p_T}  
        + 
        A \, G(x',p_T^2) \frac{d\sigma^{\,p}_{hard}}{d^2p_T}  \ ,
  \label{ATmodel}
\end{align}
where $G(x,Q^2)$ is the proton distribution function of a gluon with
fractional momentum $x$, and $x(x') \approx (p_T/\sqrt s) \exp(\pm \eta)$.
In the second term of \eqeqref{ATmodel}, 
the A-nucleus partons are assumed to undergo a
single scattering on the proton with cross section
\begin{align}
  \frac{d\sigma^{\,p}_{hard}}{d^2p_T} = \int_{x'_{min}}^1 
    \hspace*{-.35cm} dx' \ 
    G(x',p_T^2) \frac{d\hat\sigma}{d^2p_T} \ .
 \label{dsigdphard}
\end{align}
The LO pQCD cross section for gluon-gluon scattering is
$
  \frac{d\hat\sigma}{d^2p_T} 
    \approx \frac92 \alpha_s^2 \frac{1}{(p_T^2+p_0^2)^2} 
$ and the limits of integration on $x'_i$ are given by parton-parton
kinematic constraints. 
The first term of \eqeqref{ATmodel} accounts for multiple semihard
scatterings of the proton partons on 
the nucleus. The proton is considered pointlike at an impact parameter
$b$.
Nuclear effects are assumed to be due only to multiple semihard
scatterings, and are included in $d\sigma_{A}^{hard}/d^2p_T$, 
the transverse momentum distribution of a 
proton parton who suffered {\it at least} one semihard scattering.
This is written as \cite{AT01b,GV02,BFPT02}:
\begin{align}
  \frac{d\sigma^{\,A}_{hard}}{d^2p_t}
     \ = \sum_{n=1}^{\infty} \frac{1}{n!} \int d^2b \,
     \frac{d\sigma^{\,p}_{hard}}{d^2k_1} T_A(b) 
     \times \dots \times  
     \frac{d\sigma^{\,p}_{hard}}{d^2k_n} & T_A(b)
     \ e^{\, - \sigma^p_{hard}(p_0) T_A(b)} \, \nonumber \\ 
       & \times  \delta^{(2)}\big(\sum {\bf k}_i - {\bf p_t}\big) \ , 
  \label{dWdp}
\end{align}
where $\sigma_p^{hard}(p_0) = \int d^2k
\frac{d\sigma^{\,p}_{hard}}{d^2k}$ is the integrated gluon-nucleon
cross section, which depends explicitly on the infrared regulator
$p_0$. $T_A(b)$ is the target nucleus thickness function. 
The exponential factor in \eqeqref{dWdp} represents the probability
that the parton suffered no semihard scatterings after the $n$-th one. 
In such a way, unitarity is explicitly implemented at the
nuclear level, as discussed in Ref.~\cite{AT01b,BFPT02}. 
The sum over $n$ may be explicitly performed in Fourier space.
The result reads:
\begin{align}
  \frac{d \sigma^A_{hard}}{d^2p_T} = \int \frac{d^2r_T}{4\pi^2}
    e^{\,-i \vec{k}_T \cdot \vec r_T} S_{hard}^A(r_T;p_0) \ ,
 \label{spechard}
\end{align}
where 
\begin{align}
    S_{hard}^A(r_T;p_0) =\int d^2 b \, 
       \big[ e^{\,-\,\widetilde\sigma_{hard}^p(r_T,p_0)\,T_A(b)} 
         - e^{\,-\sigma_{hard}^p(p_0) T_A(b)} \big]
 \label{Shard}
\end{align}
and
\begin{align}
        \widetilde\sigma_{hard}^p(r_T;p_0) = \int d^2 k
        \left[ 1 - e^{\,-i \vec k \cdot \vec r_T} \,\right]
        \frac{d\sigma_{hard}^p}{d^2k} \ .
 \label{sighard}
\end{align}
Note that $\sigma_{hard}^p(r_T) \propto r_T^2$ as $r_T\rightarrow 0$ and
$\sigma_{hard}^p(r_T) \rightarrow \sigma_{hard}^p$ as $r_T\rightarrow
\infty$. This suggests the interpretation of $\sigma_{hard}^p(r)$ as a 
{\it dipole-nucleon ``hard''  cross section}. This dipole is of
mathematical origin, and comes from the square of the scattering
amplitude written in the Fourier variable $r_T$, which represents the
transverse size of the dipole.
Then, we can interpret $S_{hard}^A$ as the {\it dipole-nucleus ``hard''  
cross section}. \eqeqref{Shard} clearly incorporates Glauber-Gribov
multiple scatterings of the colour dipole.
Note that no nuclear effects on PDF's are included, but
shadowing is partly taken into account by the dipole multiple
scatterings.

%%%%%%%%%%%%%%%%%%%%%%%%%%%%%%%%%%%%%%%%%%%%%%%%%%%%%%%%%%%%%%%%%%%%%%%%%%
%%%%%%%%%%%%%%%%%%%%%%  SECTION 3     %%%%%%%%%%%%%%%%%%%%%%%%%%%%%%%%%%%%
%%%%%%%%%%%%%%%%%%%%%%%%%%%%%%%%%%%%%%%%%%%%%%%%%%%%%%%%%%%%%%%%%%%%%%%%%%

\section{CGC and the colour dipole}

The Colour Glass Condensate 
is an effective theory for the nucleus
gluon field at small-$x$, which describes the high-density regime
where gluon saturation comes into play to modify the free-nucleon
PDF's \cite{Iancu02}.

In this framework,
minijet production in $pA$
collisions may be related to the cross section for the scattering of a
colour dipole on a nucleus \cite{GJM02}. 
The basic assumption is that the density of
projectile partons is low enough for parton correlations inside the
proton to be due only to DGLAP evoultion. In this case one is allowed
to treat the proton in the
collinear factorization limit, exactly as in the pQCD+Glauber model of
the previus section, see \eqseqref{ATmodel} and \eqref{spechard} compared to
Eqs.~(36) and (32) of \refref{GJM02}. 

However, the dipole-nucleus interaction is computed entirely 
in the CGC model for the target nucleus.
By using the so-called ``Gaussian approximation'' \cite{Gaussapprox},  
in which parton correlation are assumed only to be
Gaussian, one may write the dipole-nucleon cross section appearing in 
\eqeqref{spechard} as follows:
\begin{align}
  \widetilde\sigma_{CGC}^p(r_T) = 4\pi\alpha_s  \, N_c  
    \int \frac{d^2k_T}{(2\pi)^2} 
    \frac{\mu_\tau(k_T)}{k_T^4} \big[ 1-e^{-ik_T \cdot r_T} \big] \ , 
 \label{SCGC}
\end{align}
where $N_c$ is the number of colours, and $\tau = \ln(1/x)$ is called
``rapidity''. Next,, $\mu_\tau$ is interpreted 
as the {\it unintegrated gluon distribution} and 
has two limits \cite{Gaussapprox}:
\begin{align}
  \mu_\tau = \left\{ \begin{array}{ll}
    \ds \frac{4\pi^2}{N_c^2-1}  
      \frac{d\, x G(x,k_T^2)}{d\, \ln(k_T^2)}
      & k_T^2 \,\rangle\rangle\, Q_s^2 \\
    \ds  \frac{\delta}{\alpha_s}  k_T^2\ln \frac{Q_s^2}{k_T^2}
      &  k_T^2 \,\langle\langle\, Q_s^2
  \end{array} \right.
 \label{sigCGC}
\end{align}
These high- and low-$k_T$ limits are defined relatively to the ``saturation
momentum'' $Q_s^2(\tau) = Q_0^2 e^{\, 4.84  \, \bar\alpha_s (\tau -\tau_0)}$, 
where $\tau_0$ is the minimum rapidity at which gluon saturation occurs and
$Q_0$ is the corresponding saturation scale. Unfortunately, the constants
$\delta$, $\tau_0$ and $Q_0$ cannot be computed in this approximation.

%%%%%%%%%%%%%%%%%%%%%%%%%%%%%%%%%%%%%%%%%%%%%%%%%%%%%%%%%%%%%%%%%%%%%%%%%%
%%%%%%%%%%%%%%%%%%%%%%  SECTION 3     %%%%%%%%%%%%%%%%%%%%%%%%%%%%%%%%%%%%
%%%%%%%%%%%%%%%%%%%%%%%%%%%%%%%%%%%%%%%%%%%%%%%%%%%%%%%%%%%%%%%%%%%%%%%%%%

\section{CGC is pQCD+Glauber}
\label{sec:CGCisGlauber}

The similarity of the CGC
model at high-$k_T$ with the pQCD+Glauber model becomes evident
by comparing \eqseqref{SCGC} and \eqref{sigCGC} with \eqseqref{Shard},
\eqref{sighard} and  \eqref{dsigdphard}.
To make this relationship more precise we proceed in four steps. 

{\bf Step 1.} In the CGC, we separate ``soft'' and ``hard'' interactions:
\begin{align}
  \mu_\tau = \mu_\tau^S + \mu_\tau^H 
  \hspace*{.4cm} {\rm with} \hspace*{.4cm}
  \left\{ \begin{array}{l}
    \ds \mu_\tau^H = \frac{4\pi^2}{N_c^2-1} \,
       \frac{d\, x G(x,k_T^2)}{d\, \ln(k_T^2)} 
       \times \frac{k_T^4}{(k_T^2+Q_s^2)^2} \\
    \ds \mu_\tau^S = \mu_\tau - \mu_\tau^H
  \end{array} \right.
 \label{step1}
\end{align}
so that $\mu_\tau^H$ vanishes quickly for $k_T<Q_s$.
Accordingly, we have a soft and hard dipole-nucleon cross section: 
$
  \ds \widetilde\sigma_{CGC}^p = \widetilde\sigma_{\rm eff}^S 
    + \widetilde\sigma_{\rm eff}^H \ . 
$

{\bf Step 2.} In the dipole-nucleus cross section, we isolate
contributions from processes with purely soft rescatterings, and
processes with at leat one hard scattering. These define,
respectively, the soft and hard dipole-nucleus cross section:
\begin{align*}
  S_{CGC}^A = & \ \ S_{\rm eff}^S \ + \ S_{\rm eff}^H 
    = \int d^2 b \, e^{\,-\, \widetilde\sigma_{\rm eff}^{S}(r_T)
    \,T_A(b)} \, e^{\,-\, \sigma_{\rm eff}^{H}
    \,T_A(b)}\\
    & + \int d^2 b \, e^{\,-\, \widetilde\sigma_{\rm eff}^S(r_T) \,T_A(b)} 
       \Big[ 
       e^{\, -\widetilde\sigma_{\rm eff}^H(r_T) \,T_A(b)} 
         - e^{\, -\sigma_{\rm eff}^H \,T_A(b)} \Big] \ .
\end{align*}
The first term includes  any number of soft scatterings but no hard
ones, and the second term processes with at least one hard scattering.

{\bf Step 3.} Study observables for which it is possible to neglect the
soft part, e.g., hadron spectra at large $p_T$ (Cronin effect), or
integrated charged multiplicities at large $\sqrt s$,  
where contribution from hard processes should become dominant. 

{\bf Step 4.} Following \refref{GM97}, we use the DGLAP equation to 
approximate the unintegrated gluon distribution in \eqeqref{step1}
with the integrated one:
$$ 
 \frac{d\, x G(x ,k_T^2)}{d\, \ln k_T^2}
    \approx  \frac{\alpha_s N_c}{\pi} 
    \int_x^1 \hspace*{-.1cm} dx'\,\,G(x',k_T^2)
$$
This approximation is valid at large $Q^2 \rangle\rangle Q_s^2$.

{\bf Result.} Collecting all the steps, we obtain 
\begin{align}
   S_{CGC}^A & \approx \int d^2b  
       \big[ e^{\,-\widetilde\sigma_{\rm eff}^H(r_T) T_A(b)} 
       - e^{\,-\sigma_{\rm eff}^H T_A(b)} \big]
 \label{result1} \\
% {\rm with} \hspace*{.6cm}
  \widetilde\sigma_{\rm eff}^H(r_T) 
    & = \int d^2k_T 
      \int_x^1 \hspace*{-.1cm} dx'\,\,G(x',k_T^2)
      \frac{(9/2)\alpha_s^2}{(k_T^2+Q_s^2)^2}
      \big[ 1-e^{-ik_T \cdot r_T} \big] \ .
 \label{result2}
\end{align}
\eqseqref{result1} and \eqref{result2} are equal to the pQCD+Glauber 
\eqseqref{Shard} and \eqref{sighard} with $p_0=Q_s$ and $x=x'_{min}$.

%%%%%%%%%%%%%%%%%%%%%%%%%%%%%%%%%%%%%%%%%%%%%%%%%%%%%%%%%%%%%%%%%%%%%%%%%%
%%%%%%%%%%%%%%%%%%%%%%  SECTION 5     %%%%%%%%%%%%%%%%%%%%%%%%%%%%%%%%%%%%
%%%%%%%%%%%%%%%%%%%%%%%%%%%%%%%%%%%%%%%%%%%%%%%%%%%%%%%%%%%%%%%%%%%%%%%%%%

\section{What do RHIC data have to say?}

Under the assumptions stated in the previous section, it is possible
to quantitatively compute some observables of interest. 
While a detailed applicaton of \eqeqref{spechard}
 to hadron $p_T$-spectra and the
Cronin effect is under investigation \cite{AG03}, we may use it to
study RHIC Au-Au data on the centrality dependence of charged particle
pseudorapidity densities \cite{Accardi01}. 
The model was derived in the case of $pA$
collisions, but is generalizable to AB collisions under the
assumption that in the regime of interest both nuclei are composed of
a diluted enough system of partons.

As a first step, we need the average minijet multiplicity at fixed impact
parameter and pseudorapidity. This is obtained by
integrating Eq.~(\ref{dWdp}) over $p_T$, and introducing the
thickness function $T_B$ of projectile nucleus:
\begin{align}
     \frac{d\Nmj}{d\eta}(b;p_0)
        =\int d^2\beta \,\, G(x,p_0^2) \,\, T_B(b-\beta) 
        \, \Bigl[1-e^{- K \, \sigma_{hard}^p(p_0) T_A(\beta)}\Bigr] 
        + A \leftrightarrow B \ ,
    \label{NmjA}
\end{align}
where the K-factor that simulates higher-order corrections to
the  pQCD cross section is explicitly shown.

\begin{figure}[t]
\begin{center}
\vskip-1.2cm 
\parbox{8cm}{\insertplot{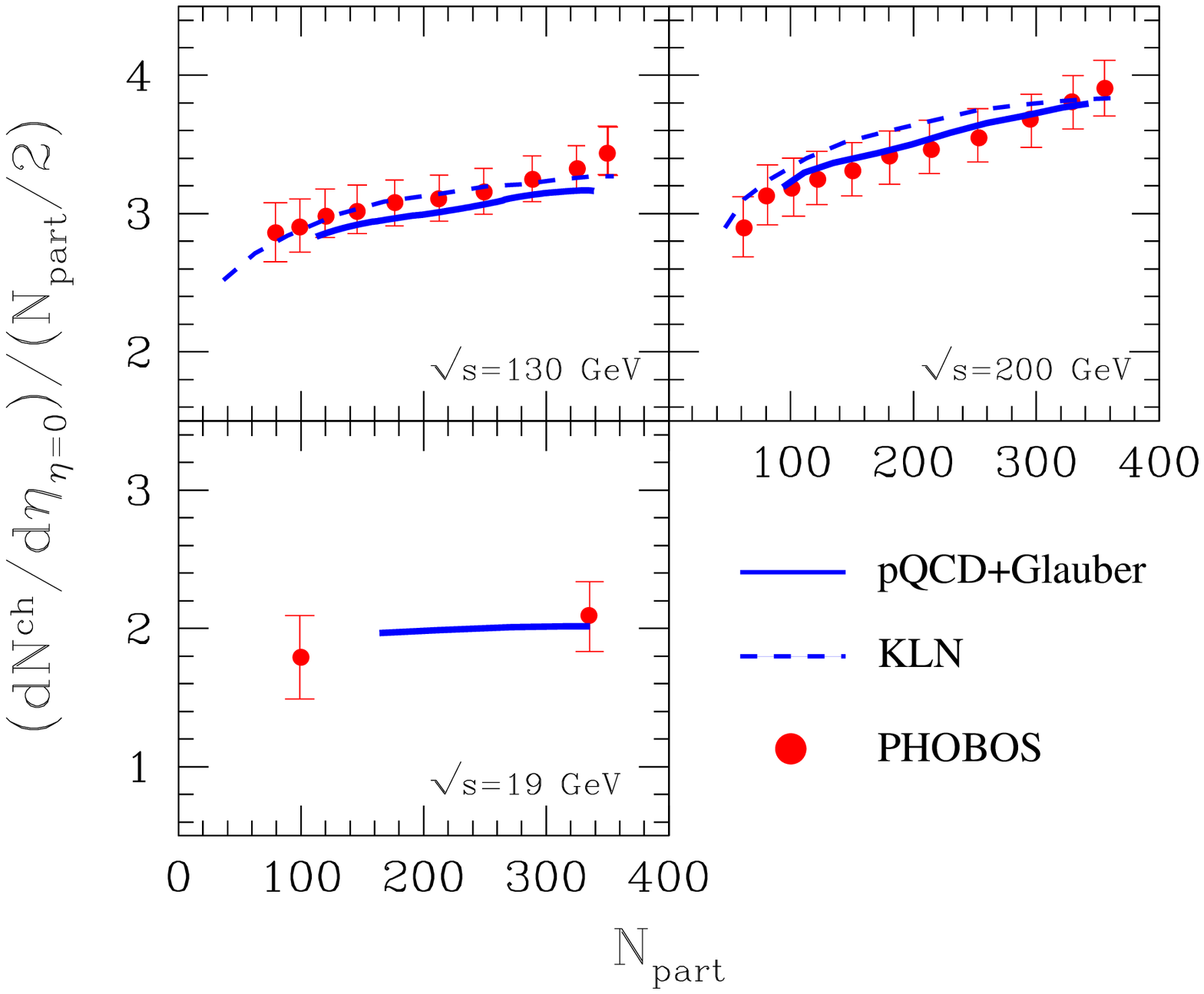} }
\parbox{4.5cm}{{\bf Fig. 1} \footnotesize
Charged particle multiplicity per participant pair at RHIC, \eqeqref{nch3}, 
as a function of the number of participants at different center of
mass energies. Solid lines are the model results for the hard plus soft
component. In the numerical computations both the gluons
and the quarks have been included. 
Dashed lines are the result of the KLN saturation model \cite{KL01}. 
Experimental data are from the PHOBOS collaboration \cite{PHOBOS02}. 
}
\vskip-1.6cm 
\end{center}
\end{figure}

The integrand in \eqeqref{NmjA} may be interpreted as the average density
of projectile partons (at a given $x$) times the probability of
having at least one semi-hard scattering against the target. 
At low values of 
$p_0$ the semihard cross section is large and the target becomes 
mor and more black to the projectile partons: the probability of scattering 
at least once becomes so high that nearly every projectile parton
scatters and is extracted from the incoming nuclear wave-function.
In this regime even if we use a lower cutoff $p_0$ no more partons 
are there to be extracted. For
this reason the minijet multiplicity tends to saturate \cite{AT01a}.
We call {\it saturation cutoff}  $p_{sat}$ the largest value of $p_0$
at which this happens, and evaluate minijet multiplicities 
from \eqeqref{NmjA} with $p_0=p_{sat}$. We refer to
\cite{Accardi01,Accardi03transp} for more details. 

To apply this minijet-level computation to integrated charged
multiplicities, we write the charged particle 
multiplicity per unit rapidity as the sum of a soft and a semi-hard
part \cite{Accardi01}. 
For the soft part we use the wounded-nucleon model, which postulates a 
scaling of $\Nch$ with the number of participants.
The semi-hard part is assumed to be completely computable from the
saturation criterion for minijet production described above.
To convert the minijet multiplicity to charged particle multiplicity,
we further assume isentropic expansion of the initially produced
minijet plasma, and parton-hadron duality.
Our final formula reads
\begin{align}
  \frac{1}{\Npart(b)/2} \frac{d\Nch}{d\eta}(b) = n_{soft} +
    \frac{1}{\Npart(b)/2}  \frac{5}{3}
                 \frac{d\Nmj}{d\eta}(b;p_{sat})  \ .
  \label{nch3}
\end{align}
As the soft component $n_{soft}=n_{soft}(\sqrt s)$ in \eqeqref{nch3} does not
depend on the centrality of the collision, we may fit it to 
data for central collision only (we used data at $\sqrt s$
= 56-200 GeV \cite{phoboscentral} and extrapolated the fit down to 19
GeV, see \refref{Accardi03transp}). 
The behaviour of the observable in non-central
collisions is then completely determined by the model.
In Fig.~1 we show the results of \eqeqref{nch3} compared
to \phobos data. Note that in our computation we used standard DGLAP
evolved PDF's from the GRV group \cite{GRV98}, with no nuclear 
modification of any sort.

%%%%%%%%%%%%%%%%%%%%%%%%%%%%%%%%%%%%%%%%%%%%%%%%%%%%%%%%%%%%%%%%%%%%%%%%%%
%%%%%%%%%%%%%%%%%%%%%%  SECTION 5     %%%%%%%%%%%%%%%%%%%%%%%%%%%%%%%%%%%%
%%%%%%%%%%%%%%%%%%%%%%%%%%%%%%%%%%%%%%%%%%%%%%%%%%%%%%%%%%%%%%%%%%%%%%%%%%

\section{Conclusions}

The pQCD+Glauber formulation gives a very well defined framework in
which detailed numerical calculations may be performed, and which is
well-tested in the case of nucleon-nucleon collisions. It is
equivalent to the CGC model in a kinematic 
region where parton densities are low enough for gluon
saturation effects to be negligible, but high enough 
for multiple scatterings and unitarity effects to become important.
RHIC data on charged multiplicities at central rapidity are very
satisfactorily described by the pQCD+Glauber model with free-nucleon PDF's.
Therefore, at least at central rapidity, 
RHIC data do not show any sign of gluon saturation.

However, integrated multiplicities have very limited
sensitivity to the details of the production processes,
and alone do not
allow any definitive conclusion. To  assess the presence or absence
of nuclear effects beyond multiscatterings and unitarity, it 
is necessary to study more differential observables 
in a cleaner environment than the hot and dense medium produced in
Au-Au collisions. Novel effects 
may then be observed as a deviation of experimental data from the
baseline given by pQCD+Glauber.
The Cronin effect on hadron production in $dA$ collisions at RHIC is an
ideal candidate.

%%%%%%%%%%%%%%%%%%%%%%%%%%%%%%%%%%%%%%%%%%%%%%%%%%%%%
%%%%%%%%%%%%%%%%%%%%%  BIBLIOGRAFIA  %%%%%%%%%%%%%%%%
%%%%%%%%%%%%%%%%%%%%%%%%%%%%%%%%%%%%%%%%%%%%%%%%%%%%%

\vfill\eject
\end{document}